\documentclass[conference, letterpaper]{IEEEtran}
\IEEEoverridecommandlockouts 

\usepackage[backend=bibtex,style=ieee]{biblatex}
\usepackage{amsmath,amssymb,amsfonts}
\usepackage{algorithm}\usepackage{algpseudocode}

\usepackage{hyperref}
\usepackage{algorithm}
\usepackage{algpseudocode}
\usepackage{graphicx,epstopdf}
\usepackage{graphics}
\usepackage{graphicx}
\usepackage{epstopdf}
\usepackage{subcaption}
\usepackage{textcomp}
\usepackage{xcolor}
\usepackage{multirow}
\usepackage{blindtext}

\graphicspath{{Figures/}}
\def\BibTeX{{\rm B\kern-.05em{\sc i\kern-.025em b}\kern-.08emT
\kern-.1667em\lower.7ex\hbox{E}\kern-.125emX}}

\begin{document}

\title{A Multi-task Learning Framework for Drone State Identification and Trajectory Prediction}

%suggested titles %
%Simultaneous Classification and Trajectory Prediction of Drones using Multi-Task Learning
%Hybrid Multi-Task Learning for Joint State and Trajectory Estimation of Drones
%Multi-Task Learning with Classification and Regression Models for Drone State Identification and Trajectory Prediction 
%Enhanced Multi-Task Learning for Accurate Drone State Identification and Trajectory Prediction 

\DeclareRobustCommand{\IEEEauthorrefmark}[1]{\smash{\textsuperscript{\footnotesize #1}}}

\author{Antreas Palamas, Nicolas Souli, Tania Panayiotou, Panayiotis Kolios, Georgios
Ellinas

\thanks{KIOS Research and Innovation Center of Excellence (KIOS CoE) and Department of Electrical and Computer Engineering, University of Cyprus. E-mail:\texttt{\{palamas.antreas, nsouli02, panayiotou.tania, kolios.panayiotis, gellinas\}} @ucy.ac.cy}}% 

\maketitle

\thispagestyle{empty}
\pagestyle{empty}

\begin{abstract}
The rise of unmanned aerial vehicle (UAV) operations, as well as the vulnerability of the UAVs' sensors, has led to the need for proper monitoring systems for detecting any abnormal behavior of the UAV. This work addresses this problem by proposing an innovative multi-task learning framework (MLF-ST) for UAV state identification and trajectory prediction, that aims to optimize the performance of both tasks simultaneously. A deep neural network with shared layers to extract features from the input data is employed, utilizing drone sensor measurements and historical trajectory information. Moreover, a novel loss function is proposed that combines the two objectives, encouraging the network to jointly learn the features that are most useful for both tasks. The proposed MLF-ST framework is evaluated on a large dataset of UAV flights, illustrating that it is able to outperform various state-of-the-art baseline techniques in terms of both state identification and trajectory prediction. The evaluation of the proposed framework, using real-world data, demonstrates that it can enable applications such as UAV-based surveillance and monitoring, while also improving the safety and efficiency of UAV operations.
\end{abstract}


\section{Introduction}
Unmanned aerial vehicles (UAVs) have gained immense popularity in recent years, attracting attention from both the research community and the industrial sector. This has resulted in the development of numerous hardware and software modules specifically designed to meet the diverse needs of the market. UAVs' customizable features make them suitable for a wide range of applications, including remote surveillance, search and rescue operations, and autonomous deliveries. However, the numerous applications of UAVs (drones) necessitate strict safety and security measures during missions. As with any other machines, drones' build-in hardware and software components are susceptible to faults and malicious cyber-attacks that may disrupt their normal operation. Thus, proper monitoring mechanisms must be in place to detect any abnormal behavior.

One of the key tasks involved in the safety assessment and flight monitoring of autonomous UAVs is to ensure that they follow a predetermined trajectory defined in the flight/mission plan within the expected time \cite{corbetta2019real}. Real-time monitoring utilizes historical trajectory as a-priori knowledge to estimate the drone's current state,  as well as to predict the remaining trajectory. The historical trajectory is updated by GPS measurements once the UAV operation is activated (i.e., the drone is in flight). The monitoring system is then responsible to compute the safety index based on the distance from obstacles, which relies heavily on the accuracy of predicting the remaining trajectory. Hence, accurately and quickly obtaining the future flight trajectory of autonomous UAVs is of great significance in reducing accidents and improving UAV operation safety levels. 

Clearly, both tasks are important for different types of drone applications. For example, trajectory prediction is vital for applications such as search and rescue, surveillance, and delivery, where accurate and reliable trajectory prediction is essential. Additionally, in terms of current state identification, this task provides the drone operator with the ability to identify the drone's state at any moment and especially in an environment where there is no visibility. Moreover, this task is useful in the event that the drone's operator perceives that the flight is being performed normally, while the drone is actually performing different movements/actions (e.g., due to a malicious attack).
 
At present, there exists a plethora of systems that attempt to identify the current state of a drone and predict its trajectory. The majority of these systems employ classification algorithms to recognize the drone's current state, while deep neural networks are utilized for trajectory prediction. For instance, certain studies have concentrated on solving multilabel classification problems to accurately identify the drone's current state \cite{georgiou2022accurate}. Alternatively, other research efforts have used deep learning models, that primarily rely on long short-term memory (LSTM) networks constructed by recurrent neural networks (RNNs), to predict the drone's trajectory for a specific number of future time-steps \cite{shu2021trajectory}.

In accordance, this work proposes a novel multi-task learning framework (MLF-ST), where the two aforementioned tasks can be performed by training and executing a single model. The proposed MLF-ST framework is trained with a common input and has the ability to output the two-fold objective of state identification and trajectory prediction. Specifically, the contributions of this work are as follows: 

\begin{itemize}
\item  A multi-task learning model is proposed that utilizes LSTM to integrate the identification of a drone's current state and the prediction of its future trajectory for a given number of time-steps. The proposed model achieves this by leveraging annotated time series data, collected from various sensors for multiple drone flights in diverse outdoor environments and conditions. This approach offers a robust solution to the complex challenge of simultaneously performing multiple tasks during the operation of the drone. 

\item A sliding window technique is implemented to segment the input data into smaller subsets, providing the model with more contextual information about the past. The approach aims to improve the accuracy of the model's predictions for the future, resulting in a robust and effective solution for data processing.

%\item The implemented framework is extensively tested on two datasets, as mentioned in Section~\ref{dataset}, in order to assess its performance in terms of current state identification and trajectory prediction under different conditions and with varying patterns of movement.
\end{itemize}

In the rest of the paper, Section~\ref{sec:relatedwork} outlines the state of the art, while the methodology and the algorithm implementation of this work are described in Sections~\ref{sec:methodology} and~\ref{sec:algorithm}, respectively. The description of the datasets and the various experiments performed and results obtained are presented in Section~\ref{sec:experiments}, while the main conclusions along with future research directions are discussed in Section~\ref{sec:conclusion}.

\section{Related Work}
\label{sec:relatedwork}
In this work, a real-world dataset is utilized \cite{rodrigues2021flight} to train and evaluate the proposed multi-task learning framework for drone state identification and trajectory prediction. The dataset includes readings from multiple sensors mounted on a DJI Matrice 100 quadcopter, which were collected during various missions with similar prescribed actions. This dataset was also employed in previous studies \cite{choudhry2021cvar,rodrigues2021drone,dai2022data} to train models for predicting UAV battery consumption. For example, in \cite{choudhry2021cvar}, temporal convolutional networks were employed  to establish an energy consumption model that predicted the energy consumption of a moving drone, in an effort to evaluate the probability of the drone depleting its energy during a pre-planned mission. Moreover, work in \cite{rodrigues2021drone} developed a machine-learning algorithm that evaluated energy usage during takeoff, cruise, and landing, utilizing distinct models for each one of the three regimes in order to predict the drone's total energy consumption during a mission. Moreover, in \cite{dai2022data}, a comprehensive and precise model for energy consumption prediction was established with the use of ensemble learning, and combining the random forest and extreme gradient boosting machine learning algorithms, demonstrating a mean absolute percentage error of approximately $10\%$.

In another comparative study of two state-of-the-art supervised approaches \cite{alos2021using}, long short-term memory (LSTM) and convolutional LSTM (ConvLSTM) were used to identify potentially defective sensors while analyzing streamed data from a UAV. The first approach involved using multiple LSTM networks to detect anomalies in the continuous values of each attribute, while the second one depended on a multi-output ConvLSTM network to identify abnormalities in the values of the concerning attribute by considering the effect of all other attributes. In addition, the authors proposed a method to address the redundancy issue that arises from using multiple LSTM networks to analyze multivariate and correlated data. A real-life dataset of four flights employing a fixed-wing aircraft was used to conduct experiments, and both approaches were able to detect different types of faults. However, it was demonstrated that the multi-output ConvLSTM is faster and achieves better results in most of the cases examined.

In terms of trajectory prediction, which is an active research topic in various fields (i.e.,  transportation, security, and robotics), a brief overview of related works that utilize multi-task learning models is provided below. For example, a recent work \cite{liu2022modeling}, proposed a multi-task learning framework for trajectory prediction, which jointly learned to predict the future locations and directions of moving objects, utilizing both spatial and temporal features to improve prediction accuracy. Moreover, the work in \cite{wu17modeling} proposed an RNN-based model for trajectory prediction (i.e., for capturing the temporal dependencies of trajectories) and used an LSTM network to model the spatial features. Also, works in \cite{qin2022typhoon,qin2022deepfr} implemented two different models for trajectory prediction; the first model was based on multimodal and multi-task learning, utilizing various weather parameters to improve the prediction of typhoon trajectories, while the second model was based on deep feature representation, which extracted features from historical trajectories using a deep neural network. Furthermore, using historical flight data \cite{shu2021trajectory} deployed an LSTM-based model to predict the future locations of UAVs. Finally, in \cite{phillips2021deep}, a deep multi-task learning framework for joint localization, perception, and prediction of moving objects was presented. Specifically, a deep neural network was utilized to learn these tasks jointly, while also improving the accuracy of trajectory prediction.

This work complements the aforementioned research efforts by implementing a novel multi-task learning framework that succeeds in {\it simultaneously} performing state classification and trajectory prediction of a UAV agent utilizing various data modalities and under different environmental conditions, aiming to create an accurate system that can be successfully deployed in emergency response situations. Further, contrary to other research efforts, the end goal of MLF-ST is to extract an accurate result for both trajectory prediction and current state identification  without utilizing two different models for the two tasks.

\section{Methodology}
\label{sec:methodology}
In this section, the methodology of the proposed framework is described in detail. At first, the framework overview provides a comprehensive description of the proposed approach, highlighting its key components and their interactions. This is followed by a brief description of LSTM and the multi-task learning techniques. Moreover, the multi-task learning section discusses the rationale behind the use of this technique in the proposed model and how it can improve the overall system performance. 

\subsection{Framework Overview}
\begin{figure}[h]
\centering
\includegraphics[width=0.5\textwidth]{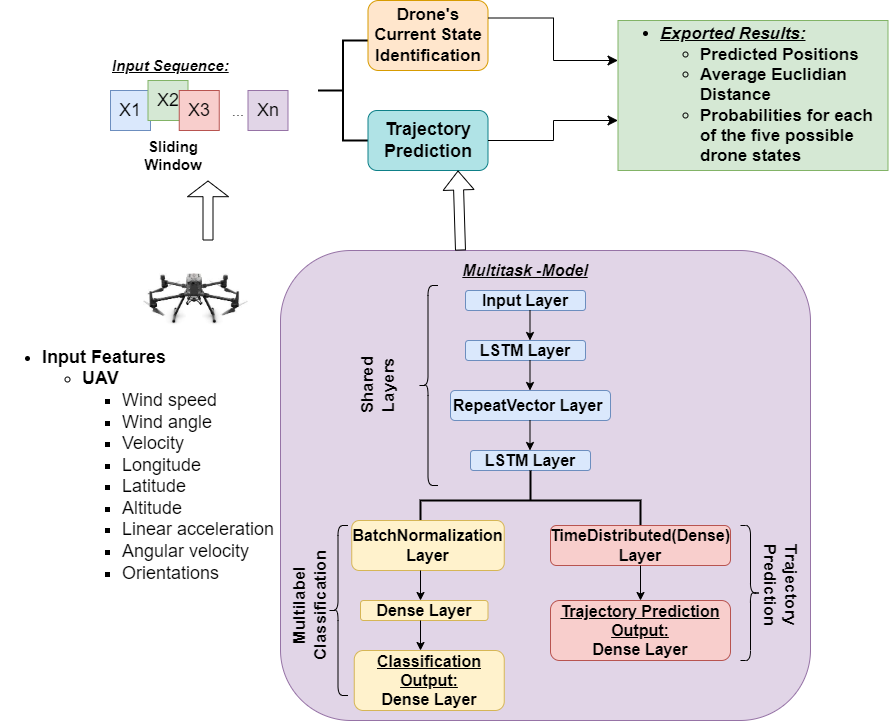}
\caption{An overview of the proposed multi-task learning framework (MLF-ST) methodology.}
\label{fig:my_label}
\end{figure}

The proposed MLF-ST framework (in Fig.~\ref{fig:my_label}) consists of a shared LSTM layer followed by separate output layers for each task. The input to the model is a window of historical drone sensor data with a fixed size ($WS$), which includes measurements such as GPS coordinates, altitude, wind speed, wind angle, velocity components of the ground speed, angular velocity components, ground acceleration, and orientation. The shared LSTM layer has $256$ units with ReLU activation, which enables the model to capture the temporal dependencies within the input data. The output of the shared  LSTM layer is then fed into a separate LSTM layer with $128$ units and ReLU activation, which is used for both tasks.

For the multilabel classification task, a batch normalization layer is applied to the output of the LSTM layer, followed by a fully connected layer with $64$ units and no activation function. The final output layer for the classification task is a sigmoid activation function that outputs a vector of probabilities for each of the possible drone states such as HOVER-IDLE, ASCENT, TURN, horizontal movement on a straight line (HMSL), and DESCENT. 

For the trajectory prediction task, a time-distributed dense layer with $64$ units is applied to the output of the biLSTM layer, which generates a sequence of predicted coordinates for each time-step in the forecast horizon (i.e., a fixed number of time-steps into the future  - typically ranging from $1$ to $3$ seconds). The final output layer for the trajectory prediction task is a linear activation function, which outputs a vector of three values representing the drone's predicted $(x,y,z)$ coordinates at each time-step in the forecast horizon. 

During the training step, a weighted combination of the mean squared error loss for the trajectory prediction task and the binary cross-entropy loss for the classification task is minimized. The Adam optimizer with a learning rate of  $0.0001$ and a batch size of $64$ is employed and the proposed model is trained for $100$ epochs. Furthermore, to circumvent overfitting, early stopping criteria is implemented, ceasing the training process when no discernible improvement is observed on the validation set. Finally, it is worth to note that exhaustive experimentation, encompassing various hidden layers and batch sizes has been conducted, with the selected hyperparameters exhibiting superior performance in terms of minimizing prediction errors.

\subsection{Background on LSTM}
LSTM is an example of an RNN architecture which is proficient in capturing long-term dependencies in sequential data by managing the flow of information through memory cells and gates \cite{hochreiter1997long}. The main building block of an LSTM network is the LSTM cell which consists of three gates ({\it input}, {\it output}, and {\it forget}) and a memory cell. In particular, the input gate controls the flow of new information into the memory cell, the output gate is responsible for the output of information from the memory cell, and the forget gate controls the retention or deletion of information from the memory cell. The gates are implemented using sigmoid functions which output values between $0$ and $1$, and the memory cell uses a hyperbolic tangent function which outputs values between $-1$ and $1$.

The equations for the LSTM cell can be expressed as:

\begin{align*}
f_t &= \sigma(W_f x_t + U_f h_{t-1} + b_f) \\
i_t &= \sigma(W_i x_t + U_i h_{t-1} + b_i) \\
\tilde{C}t &= \tanh(W_c x_t + U_c h{t-1} + b_c) \\
C_t &= f_t \odot C_{t-1} + i_t \odot \tilde{C}t \\
o_t &= \sigma(W_o x_t + U_o h{t-1} + b_o) \\
h_t &= o_t \odot \tanh(C_t)
\end{align*}

\noindent where $x_t$ is the input at time-step $t$, $h_{t-1}$ is the output from the previous time-step, $\sigma$ denotes the sigmoid function, $W_f, W_i, W_c, W_o$ are weight matrices for the forget gate, input gate, memory cell, and output gate, respectively, $U_f, U_i, U_c, U_o$ are weight matrices for the corresponding gates and memory cell from the previous time-step, and $b_f, b_i, b_o, b_c$ are bias vectors for the forget gate, input gate, output gate, and memory cell, respectively. Further, $f_t, i_t,\tilde{C}_t$ are defined as the values of the forget gate, input gate, and candidate cell state at time-step $t$, $C_t$ is the cell state at time-step $t$, $o_t$ is the value of the output gate at time-step $t$, and $h_t$ is the output at time-step $t$. Finally, $\odot$ represents element-wise multiplication. 

LSTM networks have in general achieved considerable success in diverse fields such as natural language processing \cite{sundermeyer2012lstm}, speech recognition \cite{graves2013speech}, and image captioning \cite{xu2015show}, demonstrating to be highly effective in modeling long-term dependencies in sequential data, and outperforming traditional RNNs in several tasks.  However, LSTM networks tend to be computationally expensive and require meticulous tuning of hyperparameters such as the number of LSTM cells and the learning rate \cite{bengio2015scheduled}.

\subsection{Background on Multi-task Learning}
\label{sec:training}
Multi-task learning (MTL) is a technique that aims to improve the performance of a model on multiple related tasks by jointly learning those tasks~\cite{caruana1998multitask}. In general, MTL has become increasingly popular due to its ability to improve generalization and reduce overfitting~\cite{ruder2017overview}.

MTL can be implemented using various techniques, including hard parameter sharing, soft parameter sharing, and task-specific attention \cite{ruder2017overview}. Hard parameter sharing involves sharing the same set of parameters across all tasks \cite{caruana1998multitask}, while soft parameter sharing allows for task-specific parameters that are encouraged to be similar to each other \cite{ruder2017overview}. Task-specific attention on the other hand is a technique that allows the model to selectively attend to different parts of the input for each task \cite{liu2019multi}.

MTL has been applied to different domains, including natural language processing \cite{collobert2008unified}, computer vision \cite{girshick2015fast}, speech recognition \cite{deng2013new}, and drug discovery \cite{ramsundar2015massively}. In natural language processing, MTL has been employed to jointly learn multiple tasks, such as named entity recognition and part-of-speech tagging \cite{collobert2008unified}. In computer vision, MTL is used for object detection, segmentation, and classification \cite{girshick2015fast}, while in speech recognition, is deployed to jointly learn multiple tasks, such as acoustic modeling and language modeling \cite{deng2013new}. In drug discovery, MTL has been utilized to predict multiple properties of a molecule (e.g., toxicity, solubility, etc.) \cite{ramsundar2015massively}.

%In summary, MTL is a powerful technique that can significantly improve the performance of a model on multiple related tasks. It can be implemented using various techniques, including hard parameter sharing, soft parameter sharing, and task-specific attention and can be deployed to various domains, including natural language processing, computer vision, speech recognition, and drug discovery.

\section{MLF-ST Algorithm}
\label{sec:algorithm}
This section provides a detailed description of the MLF-ST algorithm (Alg.~\ref{algo}). The MLF-ST algorithm is comprised of several stages, in an effort to enhance its functionality: (i) Stage-1: data partitioning, where a sliding window technique is employed to divide the data into subsets; (ii) Stage-2: a pre-processing phase which includes standard normalization to prepare the data for subsequent processing; (iii) Stage-3: data splitting which separates the data into distinct sets for training, testing, and validation purposes; (iv) Stage-4: the construction phase that builds the multi-task learning mode; (v) Stage-5: a compilation phase, where the model is fine-tuned for optimal efficacy; (vi) Stage-6: the fitting phase, that involves the training of the model with the use of the previously generated data subsets; (vii) Stage-7: the evaluation phase (that follows once training is finished), for assessing the accuracy and performance of the model; (viii) Stage-8: the results processing phase; this phase includes applying inverse transformations to the scaled data, obtaining original data, redefining the testing set, and making predictions. It also reshapes trajectory predictions, creates a confusion matrix for current state identification, and constructs data frames containing the actual and predicted positions ($(x,y,z)$ coordinates). Further, to refine the results, the Haversine formula is incorporated to calculate the distance between the actual and predicted positions ($x$ and $y$ coordinates), providing distance measurements in meters. The Euclidean distance formula is then employed to determine the error between the actual and predicted positions by taking into account the differences in the $z$ coordinates. 

Finally, the proposed algorithm's output is visualized, showcasing the results of both trajectory prediction and current state identification. This in-depth analysis of the MLF-ST algorithm offers valuable insights into its functioning and performance, enabling further optimization and application in various multi-task learning scenarios.

The reader should note that Alg.~\ref{algo} can also be seen as comprised of several key procedures (as shown in the pseudocode), namely the sliding window (Stage-1), model (Stages2-7), and results (Stage-8) procedures.

In the following section, experimental results are presented, demonstrating the effectiveness of the proposed multi-task framework for drone state identification and trajectory prediction.

\begin{algorithm}[ht!]
\caption{MLF-ST Algorithm}\label{algo}
\footnotesize{\textbf{Input:} {Acquired flight data}}
\begin{algorithmic}[1]
\State \footnotesize{Define input size (IS), window size (WS) and horizon size (HS)}
\Procedure \footnotesize {{Sliding Window}}{}
\State \footnotesize{Define L1, L2, L3 as empty lists}
\For{$i< IS-WS-HS+1$}
\State {\textbf{append} to L1 array Input from $i$ to WS$+i$}
\State {\textbf{append} to L2 array Input from $i+$WS to WS$+i+$HS}
\State {\textbf{append} to L3 array Input from $i+$WS to WS$+i+$HS}
\EndFor
\EndProcedure
\State\footnotesize {Reshape arrays L1 and L2 from 3D to 2D}
\State \footnotesize {Apply standard scaling on input data}
\State \footnotesize {Reshape arrays L1 and L2 from 2D to 3D}
\State \footnotesize {Split data into training, validation, and testing sets}
\Procedure \footnotesize{{Model}}{}
\State \footnotesize {Build the multi-task learning model}
\State \footnotesize {Define input and output of the model}
\State \footnotesize {Define target sets for training, testing, and validation}
\State \footnotesize {Define batch size, number of epochs, learning rate}
\State \footnotesize {Compilation of the model. Define the loss function for each task and metrics.}
\State \footnotesize {Train the model}
\State \footnotesize {Evaluate the model}
\EndProcedure
\Procedure\footnotesize {{Results}}{}
\State \footnotesize {Apply inverse transform to the scaled data}
\State \footnotesize {Get the original data}
\State \footnotesize {Define again testing set}
\State \footnotesize {Make predictions}
\State \footnotesize {Reshape trajectory predictions from 3D to 2D}
\State\footnotesize {Apply inverse transform to scaled predictions}
\State\footnotesize {Get the original values of predictions}
\State \footnotesize {Reshape again trajectory predictions from 2D to 3D}
\State \footnotesize {Make Confusion matrix for current state identification}
\State\footnotesize {Make data frames which contain  the actual and predicted positions $(x,y,z)$}
\State \footnotesize {Apply Haversine formula
 between actual and predicted positions}
\State \footnotesize {Get the distance of actual and predicted positions $(x,y)$ in meters}
\State \footnotesize {Apply Euclidean distance formula to get the error between actual and predicted positions}
\EndProcedure
\end{algorithmic}
\footnotesize{\textbf{Output:} {Trajectory prediction and current state identification results}}
\end{algorithm}

\section{Performance Evaluation}
\label{sec:experiments}
\subsection{Hardware and Software Configuration}
For the performance evaluation of the proposed MLF-ST framework a UAV is deployed in real-world experiments. Specifically, the autonomous drone is equipped with an onboard processing unit, advanced sensors, flight control systems, a solid-state weather station (TriSonica), and the NVIDIA Jetson Xavier NX Developer Kit that is responsible for saving high-precision flight data. Also, it is important to note that the data and commands are transmitted through a 4G communication architecture (i.e., a 4G dongle is applied on the moving agent). For the software implementation, the Tensorflow 2's Keras API~\cite{yang2020compositional} is used to build the multi-task learning model, while the scikit-learn library~\cite{scikit} is employed to preprocess the time series data. Also, the NumPy and Pandas Python libraries are used to analyze and visualize the results of this work.

\subsection{Dataset Collection}
\label{dataset}
This section details the sources, characteristics, and the way the data used in the experiments was annotated, including the pre-processing steps taken into account in order to ensure data quality and consistency. Specifically, in this work two different datasets are utilized, i.e., the real-world datasets provided by \cite{rodrigues2021flight} (Dataset-1) and \cite{Palamas2023-zm} (Dataset-2).

\subsubsection{Dataset-1: Data Collected with Package Delivery Quadcopter Drone}
In particular, the first dataset utilized (Dataset-1) is obtained from \cite{rodrigues2021flight}, which includes sensor data collected from a multirotor UAV during a variety of flights. Specifically, the DJI Matrice 100 (M100) UAV was used to perform autonomous takeoffs, transport payloads along a triangular flight pattern, and land. The flights were conducted at different altitudes such as $25$m, $50$m, $75$m, and $100$m, with speeds ranging from $4$m/s to $12$m/s, and payloads ranging from $0$g, $250$g, $500$g. All specific flights were repeated at least three times, for a total of $195$ flights. Specifically, this work focuses on the $182$ flights that followed the R1 route as specified in~\cite{rodrigues2021flight}, as these flights demonstrated similar flight modes at each time-step. The sensor data is synchronized at a frequency of $10$Hz and the specifics of each sensor are outlined in Table~\ref{tab1}.

\subsubsection{Dataset-2: Dataset Obtained from Multi-modal Sensor Onboard the Drone.}
This dataset contains time-series data from numerous drone flights that took place at the University of Cyprus (UCY) campus. Each flight record has a unique identifier (ID) and a timestamp indicating when the flight occurred. In particular, the drone's position is represented by the coordinates $(x,y,z)$ and altitude and the orientation of the drone is described by the quaternion $(o_x,o_y,o_z,o_w)$ (i.e., orientation x,y,z,w). The drone's linear and angular velocity are represented by $(v_x,v_y,v_z)$ and $(av_x,av_y, av_z)$, respectively. Also, the linear acceleration of the drone is defined as $(a_x,a_y,a_z)$.

Furthermore, the dataset also contains information about the battery voltage (BV), battery current (BC), and the payload attached (i.e., the devices onboard the drone, such as the Nvidia Jetson, various sensors, weather station, etc.). 

Also, the dataset includes annotations for the current state of the drone, including IDLE-HOVER, ASCEND, TURN, HMSL, and DESCEND. These states can be used for classification to identify the current state of the drone. Thus, the labeled dataset is used for predicting the trajectory of the drone using the proposed MLF-ST framework.

For the annotation procedure, the change in position $(x,y,z)$ and yaw is inspected. Specifically, if the position $(x,y)$ changes, it means that the drone moves in a horizontal straight line, and if position $z$ changes, it means that the drone performs an ascending or descending operation (depending on whether it increases or decreases). Also, if the yaw changes, it means that the drone performs a turn, and finally, if none of the above features changes, then the drone is in idle or hover mode.

In addition to the features already mentioned, the dataset also includes information from various sensors including a weather station and an inertial measurement unit (IMU). The weather station provides information about the weather conditions during the flight which includes wind speed and wind angle. These weather variables are important factors that influence the flight of the drone and battery consumption. Further, the IMU provides information on the drone's acceleration, angular velocity, and magnetic field. Specifically, the accelerometer provides information about the drone's linear acceleration, the gyroscope provides information about the drone's angular velocity, while the magnetometer measures the earth's magnetic field, which can be subsequently used to determine the drone's orientation.

Field experiments are performed in order to collect empirical data using a specific type of UAV (DJI Matrice 300 (M300)). The M300 is equipped with advanced sensors and flight control systems, providing high-precision flight data. The flights are designed  to cover a range of flight patterns, which include triangular, square, polygonal, and random flight patterns. 

\begin{table}[h]
\caption{Specifics of sensors used in Dataset-1.}
\begin{tabular}{|ll|lll}
\cline{1-2}
%\multicolumn{2}{|c|}{\textbf{TABLE   I}}             &  &  &  \\ \cline{1-2}
\multicolumn{1}{|l|}{\textbf{Features}}                        & \textbf{Summary}                         &  &  &  \\ \cline{1-2}
\multicolumn{1}{|l|}{wind speed}                     & \begin{tabular}[c]{@{}l@{}}airspeed provided by the \\ anemometer (in m/s)\end{tabular}                               &  &  &  \\ \cline{1-2}
\multicolumn{1}{|l|}{wind angle}                     & \begin{tabular}[c]{@{}l@{}}angle of the air flowing through \\ the anemometer with respect to \\ the north (in deg)\end{tabular} &  &  &  \\ \cline{1-2}
\multicolumn{1}{|l|}{battery voltage (BV)}                & \begin{tabular}[c]{@{}l@{}}system voltage measured right \\ after the battery (in V)\end{tabular}                           &  &  &  \\ \cline{1-2}
\multicolumn{1}{|l|}{battery current (BC)}               & \begin{tabular}[c]{@{}l@{}}system current measured right \\ after the battery (in A)\end{tabular}                          &  &  &  \\ \cline{1-2}
\multicolumn{1}{|l|}{position $(x,y)$}     & \begin{tabular}[c]{@{}l@{}}longitude \& latitude of\\ the UAV (in deg)\end{tabular}      &  &  &  \\ \cline{1-2}
\multicolumn{1}{|l|}{position $z$}                    & \begin{tabular}[c]{@{}l@{}}altitude of the UAV with \\ respect to the sea-level (in m)\end{tabular}                         &  &  &  \\ \cline{1-2}
\multicolumn{1}{|l|} {orientation $(o_x,o_y,o_z,o_w)$} & UAV orientation (in quartenions) &  &  &  \\ \cline{1-2}
\multicolumn{1}{|l|}{velocity $(v_x,v_y,v_z)$}            & \begin{tabular}[c]{@{}l@{}}velocity components of ground\\ speed (in m/s)\end{tabular}                                &  &  &  \\ \cline{1-2}
\multicolumn{1}{|l|}{angular velocity $(av_x,av_y,av_z)$}              & \begin{tabular}[c]{@{}l@{}}angular velocity components \\ (in rad/s)\end{tabular}                                    &  &  &  \\ \cline{1-2}
\multicolumn{1}{|l|}{linear acceleration $(a_x,a_y,a_z)$} & \begin{tabular}[c]{@{}l@{}}ground acceleration in (m/s\textasciicircum{}2)\end{tabular}                    &  &  &  \\ \cline{1-2}
\multicolumn{1}{|l|}{payload}                         & \begin{tabular}[c]{@{}l@{}}mass of the payload \\ attached to the UAV (in g)\end{tabular}                                            &  &  &  \\ \cline{1-2}
\end{tabular}
\label{tab1}
\end{table}

\subsection{Experimental Results}
This section showcases the experimental results for the proposed MLF-ST framework by comparing the localization performance for each estimated trajectory with the ground truth (i.e., measurements acquired via GPS+IMU), while also demonstrating the current state identification by performing multilabel classification for each aforementioned dataset.

To investigate the average performance of the proposed MLF-ST framework concerning Dataset-2, $10$ real outdoor experiments are conducted in the area of the UCY campus. For the purposes of the experiments, the M300 drone is utilized  (specifications about the UAV agent utilized can be found in {\it https://www.dji.com/matrice-300}).

\begin{table}[H]
\caption{Average Euclidean prediction error for Dataset-1.}
\begin{tabular}{|c|c|c|}
\hline
\multirow{4}{*}{All testing flights} & Prediction time [s] & Euclidean distance error [m] \\ \cline{2-3} 
& t+1                 & 2.5521                       \\ \cline{2-3} 
& t+2                 & 2.6637                       \\ \cline{2-3} 
& t+3                 & 3.4412                       \\ \hline
\end{tabular}
\label{tab2}
\end{table}

\subsubsection{Trajectory Prediction - Dataset-1}
Table~\ref{tab2} tabulates the location estimation accuracy of the proposed MLF-ST framework for Dataset-1. Even though the estimated location performance decreases, as the time-step ($t$) increases, the proposed system is able to provide accurate location estimation that is comparable to the ground truth values (with Euclidean distance error lower than $3.5$m).

\begin{figure}[H]
\centering
\includegraphics[width = 9.5cm, height = 3cm]{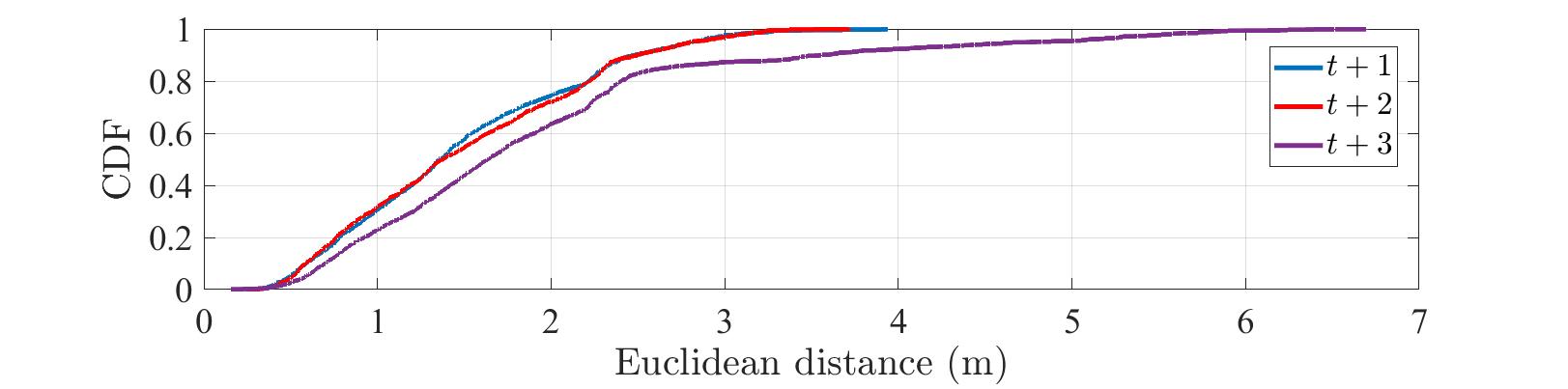}
\caption{Average error CDF for predicted trajectories at $t+1$, $t+2$, $t+3$ (Dataset-1).}
\label{fig:CDF}
\end{figure}

Further, Fig. \ref{fig:CDF} illustrates the cumulative distribution function (CDF) of the forecast for time intervals $t+1$, $t+2$, and $t+3$. The analysis of the CDF for prediction errors highlights that both the mean prediction errors and the overall dispersion of errors are indeed considerably small.

Specifically, $90\%$ of the $1$-second look-ahead predictions manifest an error magnitude less than $2.45$ meters, while the $2$-second look-ahead accounts for $90\%$ of errors falling below $2.5$ meters. Also, the $3$-second horizon encompasses $90\%$ of errors not exceeding $3.56$ meters.

\subsubsection{Current State Identification Results (Dataset-1)}
\begin{figure}[H]
\centering
\includegraphics[width = 0.5\textwidth]{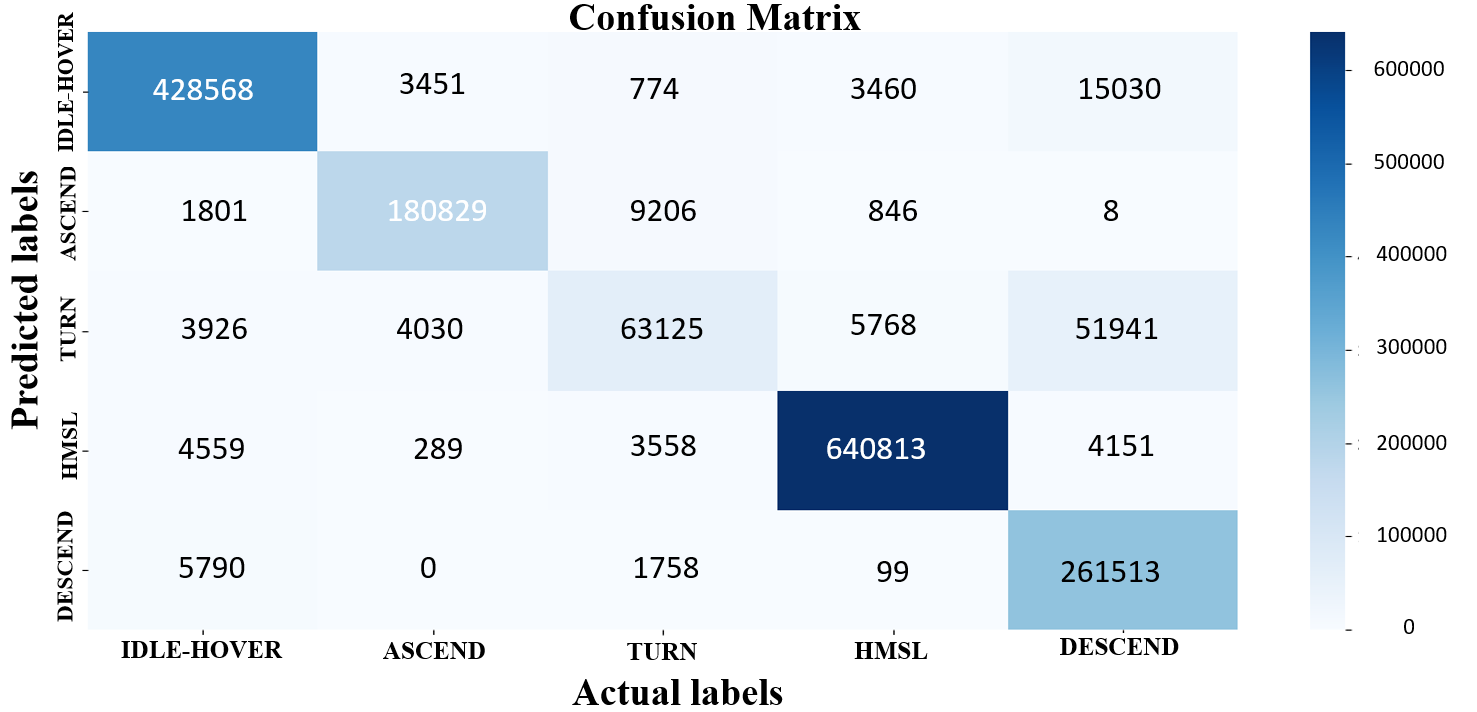}
\caption{Multilabel classification confusion matrix.}
\label{fig:my_cm}
\end{figure}

In this work, regarding the second main task (current state identification), the performance analysis of the multilabel classification for the drone's movement is presented with the use of a confusion matrix. Dataset-1 is comprised of five classes, namely IDLE-HOVER, ASCEND, TURN, HMSL, and DESCEND. The confusion matrix, as shown in Fig.~\ref{fig:my_cm}, represents the model's performance in classifying these movements.

To better understand the performance of the proposed framework's multilabel classification, the Precision is computed. Moreover, the Recall, and F1-score for each class are presented in Table~\ref{tab3}. To calculate the Precision, Recall, and F1-score metrics, the following equations can be used:

\begin{equation}
Precision = \frac{TP}{TP + FP}
\end{equation}

\begin{equation}
Recall = \frac{TP}{TP + FN}
\end{equation}

\begin{equation}
F1-score = \frac{2 \cdot (Precision \cdot Recall)}{Precision + Recall}
\end{equation}

The results indicate that the model performs considerably well for the IDLE-HOVER, ASCEND, and HMSL classes, with F1-scores greater than $0.93$. However, the model struggles with the TURN and DESCEND classes, with F1-scores of $0.5898$ and $0.8836$, respectively.

\begin{table}[h]
\caption{Class-wise metrics.}
\centering
\begin{tabular}{llll}
%\multicolumn{4}{c}{\textit{TABLE III: Class-wise metrics}}                                            \\ \hline
\multicolumn{1}{|l|}{Class}      & \multicolumn{1}{l|}{Precision} & \multicolumn{1}{l|}{Recall} & \multicolumn{1}{l|}{F1-score} \\ \hline
\multicolumn{1}{|l|}{IDLE-HOVER} & \multicolumn{1}{l|}{0.9671}    & \multicolumn{1}{l|}{0.9333} & \multicolumn{1}{l|}{0.9499}   \\ \hline
\multicolumn{1}{|l|}{ASCEND}     & \multicolumn{1}{l|}{0.9561}    & \multicolumn{1}{l|}{0.9186} & \multicolumn{1}{l|}{0.937}    \\ \hline
\multicolumn{1}{|l|}{TURN}       & \multicolumn{1}{l|}{0.795}     & \multicolumn{1}{l|}{0.4683} & \multicolumn{1}{l|}{0.5898}   \\ \hline
\multicolumn{1}{|l|}{HMSL}       & \multicolumn{1}{l|}{0.9842}    & \multicolumn{1}{l|}{0.9826} & \multicolumn{1}{l|}{0.9834}   \\ \hline
\multicolumn{1}{|l|}{DESCEND}    & \multicolumn{1}{l|}{0.8079}    & \multicolumn{1}{l|}{0.975}  & \multicolumn{1}{l|}{0.8836}   \\ \hline
\end{tabular}
\label{tab3}
\end{table}

Also, for the objective of conducting a multilabel classification analysis, overall pertinent micro-average and macro-average metrics are meticulously computed and evaluated for Precision, Recall, and F1-score. The micro-average metrics are higher than the macro-average metrics as shown in Table~\ref{tab4}, indicating high overall model performance. However, the macro-average metrics present that there is still room for improvement in the classification performance, particularly for the TURN and DESCEND classes.

\begin{table}[h]
\caption{Average metrics.}
\centering
\begin{tabular}{lll}
%\multicolumn{3}{c}{\textit{TABLE IV: Average metrics}}                                            \\ \hline
\multicolumn{1}{|l|}{Metric}    & \multicolumn{1}{l|}{Micro-Average} & \multicolumn{1}{l|}{Macro-Average} \\ \hline
\multicolumn{1}{|l|}{Precision} & \multicolumn{1}{l|}{0.9877}        & \multicolumn{1}{l|}{0.9021}    \\ \hline
\multicolumn{1}{|l|}{Recall}    & \multicolumn{1}{l|}{0.9825}        & \multicolumn{1}{l|}{0.8556}    \\ \hline
\multicolumn{1}{|l|}{F1-score}  & \multicolumn{1}{l|}{0.9851}        & \multicolumn{1}{l|}{0.8687}    \\ \hline
\end{tabular}
\label{tab4}
\end{table}

In summary, the multilabel classification task is investigated, exhibiting considerable efficacy in discerning various drone movement categories. However, it is imperative to address the sub-optimal performance observed in the TURN and DESCEND classifications. Potential strategies for refinement may encompass the incorporation of supplementary training data, the application of feature engineering techniques, or the exploration of alternative algorithmic approaches. Subsequent research endeavors will concentrate on these facets with the intent of augmenting the model's performance, thereby contributing to the evolution of robust and precise drone movement classification systems.

\subsubsection{Trajectory Prediction Results  (Dataset-2)}
Table~\ref{tab5} presents the spatial estimation precision of the proposed MLF-ST framework when Dataset-2 is utilized. Despite the decline in the estimated location performance as the temporal interval ($t$) expands, the proposed framework delivers precise spatial approximations that are analogous to the ground truth measurements (exhibiting Euclidean distance error less than $8.1$ meters).

\begin{table}[H]
\caption{Average Euclidean prediction error (Dataset-2).}
\begin{tabular}{|c|c|c|}
\hline
\multirow{4}{*}{All testing flights} & Prediction time [s] & Euclidean distance error [m] \\ \cline{2-3} 
& t+1                 & 6.8166                       \\ \cline{2-3} 
& t+2                 & 7.1228                       \\ \cline{2-3} 
& t+3                 & 8.0021                       \\ \hline
\end{tabular}
\label{tab5}
\end{table}

\begin{figure}[H]
\centering
\includegraphics[width = 9.5cm, height = 3cm]{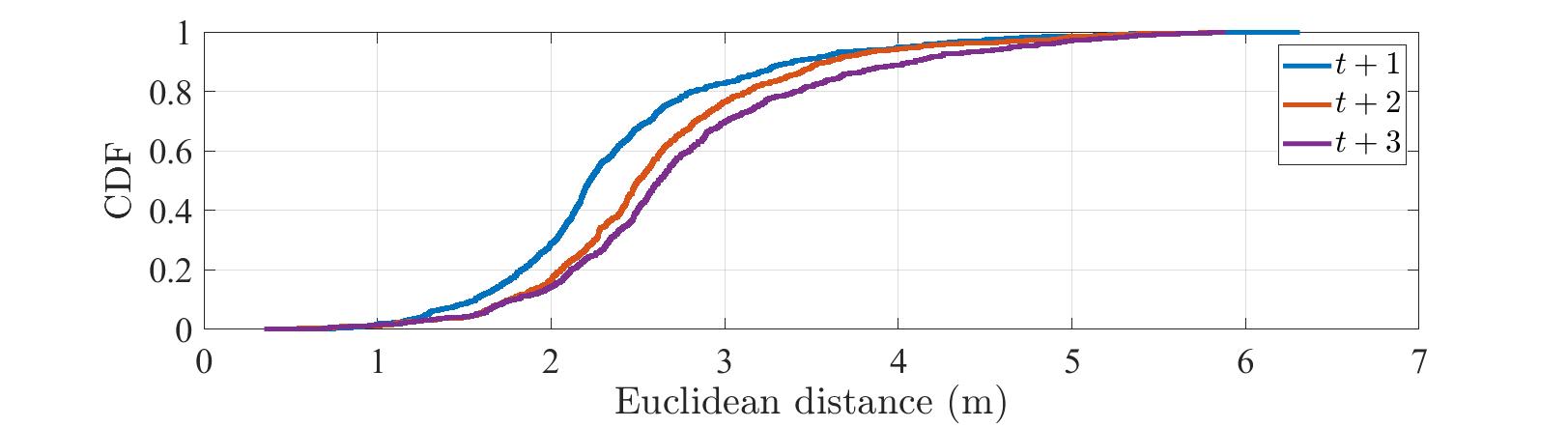}
\caption{Average error CDF for predicted trajectories at $t+1$, $t+2$, $t+3$ (Dataset-2).}
\label{fig:our_CDF}
\end{figure}

Figure \ref{fig:our_CDF} depicts the CDF of the trajectory predictions for the time intervals $t+1$, $t+2$, and $t+3$. It must be noted that the empirical data employed for the extraction of the distinct CDFs encompassed an array of motion patterns, including triangular, square, circular, and stochastic trajectories. In particular, for the $1$-second look-ahead predictions, $90\%$ exhibit an error magnitude below $3.4$ meters. Also, the $2$-second look-ahead accounts for $90\%$ of errors that are under $3.59$ meters, while for the $3$-second time horizon, $90\%$ of the errors do not surpass $4.2$ meters.

It is worth mentioning that Dataset-1  comprises of $227,552$ entries, whereas Dataset-2 contains only $95,840$ entries, leading to a lower trajectory prediction performance. The small number of entries for Dataset-2 is also the reason for not utilizing it for state identification purposes. Nevertheless, it is important for the reader to note that the expansion of Dataset-2 can potentially lead to a better or analogous system performance to the results obtained with Dataset-1. This is currently in progress, as Dataset-2 is continuously being augmented with new data points obtained through various drone flights.  

In this study, MLF-ST framework resulted in average Euclidean distance errors of 2.5521m, 2.6637m, and 3.4412m for the first three seconds using a sliding window technique with WS equals 30 and HS equals 30, respectively. In comparison, another study \cite{shu2021trajectory} reported an overall mean error of 1.5m. It is essential to consider potential differences in methodologies and datasets when comparing these results. It is worth mentioning that, in study \cite{shu2021trajectory}, the mean error was extracted after four flights, compared to the specific work that the results were extracted using a dataset equipped from 195 flights.

Regarding the multilabel classification task, the MLF-ST framework in this study achieved an accuracy of $98.51\%$, closely matching the compared paper's $98.5\%$ accuracy \cite{georgiou2022accurate}. This demonstrates competitive performance in classification accuracy. 

\section{Conclusion}
\label{sec:conclusion}
A novel multi-task learning framework (MLF-ST) is presented, leveraging deep neural networks to simultaneously optimize the desired two-fold objective of drone state identification and trajectory prediction. The proposed framework, that utilizes shared layers and various metrics to extract information from a set of input data, can efficiently learn the useful features for both tasks. Further, by incorporating LSTM networks and sliding window techniques along with the drone's historical data, MLF-ST is able to accurately identify the drone's current state and predict its future trajectory. An extensive evaluation of the presented framework showcases its ability to contribute to the overall safety and efficiency of drone operations. Also, the extensive testing conducted on two different datasets, collected under various flight conditions, further validates the robustness and adaptability of the MLF-ST framework.

Future research avenues include the extension of the proposed MLF-ST framework by incorporating additional tasks, different deep-learning techniques, and the employment of various sensor modalities to enhance the capabilities of drone monitoring systems.

\section*{Acknowledgements}\nonumber
This work was supported by the Secure and Safe Multi-Robot Systems (SESAME) H2020 Project under Grant Agreement 101017258. It was also supported by the European Union's Horizon 2020 research and innovation program under grant agreement No 739551 (KIOS CoE - TEAMING) and from the Republic of Cyprus through the Deputy Ministry of Research, Innovation and Digital Policy.

\printbibliography

\end{document}